\documentclass[a4paper,12pt,pdf]{article}
\usepackage{amsfonts}

\usepackage[dvips]{graphicx}
\usepackage{amsmath}
\usepackage{multicol}

\input epsf
\newcommand{\be}{\begin{equation}}
\newcommand{\ee}{\end{equation}}
\newcommand{\bea}{\begin{eqnarray}}
\newcommand{\eea}{\end{eqnarray}}
\newcommand{\ba}{\begin{array}}
\newcommand{\ea}{\end{array}}

\renewcommand{\Im}{{\tt Im\,}}
\renewcommand{\Re}{{\tt Re\,}}

\begin{document}

\begin{titlepage}

    \thispagestyle{empty}
    \begin{flushright}
        \hfill{CERN-PH-TH/092}\\
        \hfill{UCB-PTH-08/76}\\
        \hfill{SU-ITP-08/29}\\
    \end{flushright}

\vspace{15pt}
\begin{center}
{\LARGE{\textbf{Quantum Lift of Non-BPS Flat Directions}}}\\
\vspace{50pt}
{{\textbf{S. Bellucci$^1$, S. Ferrara$^{2,1,3}$, A. Marrani$^{4,1}$ and A. Shcherbakov$^{5,1}$}}}\\
\vspace{15pt}
{${}^1$ \it INFN - Laboratori Nazionali di Frascati, \\
Via Enrico Fermi 40,00044 Frascati, Italy\\
\texttt{bellucci, marrani, ashcherb@lnf.infn.it}}

\vspace{10pt}
{${}^2$ \it Physics Department, Theory Unit, CERN, \\
CH 1211, Geneva 23, Switzerland\\
\texttt{sergio.ferrara@cern.ch}}

\vspace{10pt}
{${}^3$ \it Miller Institute for Basic Research in Science,\\
        University of California, Berkeley, CA94720, USA}

\vspace{10pt}
{${}^4$ \it Stanford Institute for Theoretical Physics\\
Stanford University, Stanford, CA 94305-4060, USA}

\vspace{10pt}
{${}^5$ \it Museo Storico della Fisica e\\
        Centro Studi e Ricerche ``Enrico Fermi"\\
        Via Panisperna 89A, 00184 Roma, Italy}

\end{center}

\begin{abstract}
We study~$N=2$,~$d=4$ attractor equations for the quantum corrected
two-moduli prepotential~$\mathcal{F}=st^2+i\lambda$,
with~$\lambda$ real, which is the only correction which preserves
the axion shift symmetry and modifies the geometry.

In the classical case the black hole effective potential is known
to have a flat direction. We found that in the presence of~$D0-D6$
branes the black hole potential exhibits a flat direction in the
quantum case as well. It corresponds to non-BPS~$Z\neq 0$
solutions to the attractor equations. Unlike the classical case,
the solutions acquire non-zero values of the axion field.

For the cases of~$D0-D4$ and~$D2-D6$ branes the classical flat
direction reduces to separate critical points which turn out to
have a vanishing axion field.

\end{abstract}

\end{titlepage}
\newpage 

\section{Introduction}

\label{Intro}

The attractor mechanism was firstly described in the seminal papers~\cite
{FKS}-\cite{FGK} and is now the object of intense studies (for a
comprehensive list of references, see \textit{e.g.} \cite{bellucci2}). While
originally this mechanism was discovered in the context of extremal BPS
black holes, later it was found to be present even for non-BPS ones.
Differently from the~BPS black holes, such new attractors do not saturate
the BPS bound and thus, when considering a supergravity theory, they break
all supersymmetries at the black hole event horizon~\cite{BPS}.

Attractor mechanism equations are given by the condition of extremality~\cite
{FGK}
\begin{equation}  \label{Mon-1}
\phi _{H}\left( p,q \right):\qquad \left. \frac{\partial V_{BH}\left( \phi,
p ,q \right)}{\partial\phi^a}\right| _{\phi =\phi _{H}\left( p,\,q \right)}=0
\end{equation}
of the so-called black hole potential~$V_{BH}$, which is a real function of
the moduli~$\phi^a$ and magnetic~$p^{\Lambda}$ and electric~$q_{\Lambda}$
charges.

The crucial condition for a critical point~$\phi _{H}\left(p,q\right)$ to be
an attractor in the strict sense is that the Hessian matrix
\begin{equation}  \label{Mon-2}
\mathcal{H}_{ab}\left( p, q \right) = \nabla_a \nabla_b V_{BH}%
\rule[-0.5em]{0.4pt}{1.5em}_{\;\phi=\phi_H} = \partial_a \partial_b V_{BH}%
\rule[-0.5em]{0.4pt}{1.5em}_{\;\phi=\phi_H}
\end{equation}
of $V_{BH}$ evaluated at the critical point~(\ref{Mon-1}) be positive
definite.

In~$N=2$, $d=4$ Maxwell-Einstein supergravities based on homogeneous scalar
manifolds, the Hessian matrix has in general either positive or zero
eigenvalues. The latter ones correspond to massless Hessian modes, which
have been proven to be flat directions of~$V_{BH}$~\cite
{Ferrara-Marrani-1,ferrara4}.

The presence of flat directions does not contradict the essence of the
attractor mechanism: although the moduli might not be stabilized, the value
of the entropy does not change when the moduli change along the flat
directions of~$V_{BH}$. Indeed, in~$N=2$~$d=4$ supergravity, the black hole
entropy is related to its potential through the formula~\cite{FGK}
\begin{equation}
S_{BH}\left( p,q\right) =\pi \,V_{BH}\left( \phi ,p,q\right)
\rule[-0.5em]{0.4pt}{1.5em}_{\;\phi =\phi _{H}}.  \label{SBH}
\end{equation}
Therefore, whether the flat directions are present or not, it does not
affect the value of the entropy. Consequently, one may allow the eigenvalues
of the Hessian matrix to be zero, as well.

Actually, this phenomenon always occurs in $N>2$-extended, $d=4$
supergravities, also for $\frac{1}{N}$-BPS configurations, and it can be
understood through an $N=2$ analysis, as being due to $N=2$ hypermultiplets
always present in these theories \cite{Ferrara-Marrani-1,bellucci2}.

In $N=2$, $d=4$ supergravity with more than one vector multiplet coupled to
the supergravity one, the black hole potential~$V_{BH}$ has flat directions
provided that the critical points exist~\cite{ferrara4,bellucci1}. They
correspond to non-BPS states with non-vanishing central charge.

The simplest model possessing a flat direction is that with two vector
multiplets, i.e. the so-called~$st^2$ model. The latter we treat in this
paper which might be thought of as a continuation of the investigation
started in an earlier paper~\cite{BFMS1}, where we found an effect of
multiplicity of the attractors in the presence of quantum corrections. This
effect is related to the fact that when quantum corrections are introduced,
the scalar manifold is not simply connected any more.

Even in the classical case, solutions for the attractor equations are known
just for quite a few models. For example, in the framework of special
K\"ahler d-geometry, supersymmetric attractor equations are solved in~\cite
{Shmakova}. Non-supersymmetric ones are solved completely both for the~$t^3$
model~\cite{Saraikin-Vafa-1} and for the~$stu$ one~\cite{stu-unveiled},
taking advantage of the presence of a large duality symmetry. States with
vanishing central charge are investigated in~\cite{BMOS-1,stu-unveiled}.

As it has been already mentioned, in the paper~\cite{BFMS1} we began the
study of a quantum~$t^{3}$ model of~$N=2$~$d=4$ supergravity with the
prepotential\footnote{%
In general, $\lambda $ is related to perturbative quantum corrections at the
level of non-linear sigma model, computed by $2$-dimensional CFT techniques
on the world-sheet. For instance, in Type $IIA$ $CY_{3}$-compactifications
\cite{CDLOGP1,HKTY,Quantum-N=2}
\begin{equation*}
\lambda =-\frac{\chi \zeta \left( 3\right) }{16\pi ^{3}},
\end{equation*}
where $\chi $ is the Euler character of $CY_{3}$, and $\zeta $ is the
Riemann zeta-function. Within such a framework, it has been shown that $%
\lambda $ has a $4$-loop origin in the non-linear sigma-model \cite
{Alvarez-Gaume, Grisaru, CDLOGP1}.}
\begin{equation*}
F(X)=\frac{(X^{1})^{3}}{X^{0}}+i\lambda (X^{0})^{2}=(X^{0})^{2}\left(
t^{3}+i\lambda \right) ,\qquad \lambda \in \mathbb{R}.
\end{equation*}
There it was argued that this is the only possible correction preserving the
axion shift symmetry and that it cannot be reabsorbed by a field
redefinition~\cite{Peccei-Quinn,CFG}. The black hole potential of this model
does not possess any flat direction, nevertheless, the appearance of the
quantum contribution reveals an effect of multiplicity of the attractors.
This effect is similar to that observed in~\cite{G}. Due to this effect
other ones arise such as~``transmutations'' and ``separation'' of
attractors. In~$st^{2}$ model they appear as well, but here we are mostly
concerned with another phenomenon, not present in~$t^{3}$ model,~-- namely,
how the flat direction of the ~$st^{2}$ model undergoes the insertion of
quantum corrections.

The quantum corrected~$st^{2}$ model that we consider is based on the
holomorphic prepotential
\begin{equation*}
F(X)=\frac{X^{1}\left( X^{2}\right) ^{2}}{X^{0}}+i\lambda
(X^{0})^{2}=(X^{0})^{2}\left( st^{2}+i\lambda \right) ,\qquad \lambda \in
\mathbb{R}.
\end{equation*}
The complex moduli~$s$ and~$t$ span the rank-$2$ special Kähler manifold~%
$\left( SU\left( 1,1\right) /U\left( 1\right) \right) ^{2}$. When~$\lambda
=0 $ this formula gives classical expression for the prepotential, which we
start the next section with.

Knowing the superpotential, one may easily calculate the corresponding black
hole potential\footnote{%
Generally, the indices~$a,b,c,\ldots $ run from 1 to~$n$, while~$\Lambda
,\Sigma ,\ldots $~-- from 0 to~$n$, with~$n=2$ for the~$st^{2}$ model}~\cite
{FGK}
\begin{equation}
V_{BH}=e^{K}\left[ W\bar{W}+g^{a\bar{b}}\nabla _{a}W\bar{\nabla}_{\bar{b}}%
\bar{W}\right]  \label{VBH}
\end{equation}
in terms of the superpotential~$W$ and the Kähler potential~$K$
\begin{equation}
W=q_{\Lambda }X^{\Lambda }+p^{\Lambda }F_{\Lambda },\qquad K=-\ln \left[
-i\left( X^{\Lambda }\bar{F}_{\Lambda }-\bar{X}^{\Lambda }F_{\Lambda
}\right) \right] .  \label{WK}
\end{equation}

\section{$D0-D4$ branes}

\label{Sect2} This brane configuration corresponds to vanishing charges~$q_a$
and~$p^0$. The quartic invariant in this case is given by
\begin{equation}
I_4= 4 q_0 p^1 \left( p^2\right)^2.
\end{equation}
When it is negative, the classical black hole potential possesses a
non-compact flat direction related to the~$SO\left( 1,1\right)$ manifold~
\cite{ferrara4}
\begin{equation}  \label{flatDir1}
\begin{array}{l}
\displaystyle\Im s = \pm\,\sqrt{-\frac{p^1 q_0}{(p^2)^2}}\; \frac{%
\displaystyle (\Re t)^2 + \frac{q_0}{p^1}}{\displaystyle(\Re t)^2 - \frac{q_0%
}{p^1}}\,, \qquad \Re s = \frac{p^1 q_0}{p^2}\, \frac{2\Re t}{\displaystyle%
(\Re t)^2 - \frac{q_0}{p^1}}\,, \\
\displaystyle \Im t = \pm \sqrt{-\frac{q_0}{p^1}-(\Re t)^2}
\end{array}
\end{equation}
parameterized, for instance, by the real part of the modulus~$t$. Naturally,
it solves the criticality condition of the black hole potential~(\ref{VBH})
evaluated when~$\lambda=0$
\begin{equation}  \label{VBHcrit}
\frac{\partial V_{BH}}{\partial s} = 0, \qquad \frac{\partial V_{BH}}{%
\partial t} = 0
\end{equation}
and corresponds to a non-BPS state. The black hole entropy~(\ref{SBH}) turns
out not to depend on~$\Re t$
\begin{equation}
S_{BH} = \pi \sqrt{-I_4}=2 \pi \sqrt{-q_0p^1\left( p^2\right)^2}
\end{equation}
in complete agreement with the attractor mechanism paradigm.

When switching the quantum parameter~$\lambda$ on, it is convenient to pass
to the rescaled moduli~$y^1,y^2$ and the quantum parameter~$\alpha$
\begin{equation}  \label{Lun-1}
s = p^1\sqrt{- \frac{q_0}{p^1(p^2)^2}}\,y^1,\quad t = p^2\sqrt{- \frac{q_0}{%
p^1(p^2)^2}}\,y^2,\quad \lambda = q_0\sqrt{-\frac{q_0}{p^1(p^2)^2}}\;\alpha\,
\end{equation}
in order to factorize the dependence of~$W$ and~$V_{BH}$ on the charges
\begin{equation}  \label{WVfactorized}
W = q_0 \left[\rule{0pt}{1em} 1 - 2 y^1 y^2 - (y^2)^2 \right], \qquad V_{BH}
= \frac12\,\sqrt{-I_4}\;v(y,\bar y) = \sqrt{\strut -q_0 p^1 (p^2)^2}\;
v(y,\bar y).
\end{equation}
The expression for the black hole potential is quite cumbersome and not too
illustrative, so we restricted ourselves to writing down explicitly only the
superpotential. The function~$v(y,\bar y)$ is a rational one with the
numerator being a polynomial of ninth degree and the denominator~-- of
eighth degree on~$y^a$ and~$\bar y^a$. So at the moment it is quite
improbable to resolve attractor mechanism equations~(\ref{VBHcrit})
analytically. Nevertheless, numerical simulations show that all solutions to
eqs.~(\ref{VBHcrit}) have vanishing values of the axion fields
\begin{equation}  \label{axion-free}
\Re y^1=\Re y^2=0.
\end{equation}
This result differs from that present in the classical case~(\ref{flatDir1}%
). With this assumption, the attractor mechanism equations become
\begin{equation}  \label{crit-axion-free}
\begin{array}{l}
\displaystyle 4 \alpha^4 - \alpha^3 \left[ \rule{0pt}{1em} -4t_1^2 t_2 -
2t_2 (-3+t_2^2) + 2 t_1 (3 + t_2^2) \right] + \alpha^2 t_1 t_2 \left[ \rule%
{0pt}{1em} 5 + 32 t_1^2 t_2^2 + 11 t_2^4+ \right. \\
\displaystyle\phantom{4 \alpha^4+} \left. + t_1 (-6 t_2 + 26 t_2^3) \rule%
{0pt}{1em}\right] -4\alpha t_1^2t_2^3\left[ \rule{0pt}{1em} -1 + 3 t_2^2 + 2
t_1^2 t_2^2 + 2 t_2^4 + t_1 t_2 (9 + t_2^2)\right] + \\
\displaystyle \phantom{4 \alpha^4+} - 8 t_1^3 t_2^5 \left( -1 + t_2^4
\right)=0, \\[0.5em]
\displaystyle 4 \alpha^4 + 4 \alpha^3 t_2 (-3 + t_2^2) + \alpha^2 t_2^2
\left[ \rule{0pt}{1em} 5 + (-6 + 32 t_1^2) t_2^2 + 32 t_1 t_2^3 + 5 t_2^4%
\right] - \\
\displaystyle \phantom{4 \alpha^4+} -4\alpha t_1 t_2^4 \left[ \rule{0pt}{1em}
-1 + 6t_2^2 + 4t_1^2 t_2^2 - t_2^4 + 2 t_1 t_2 ( 3 + t_2^2)\right] + 8 t_1^2
t_2^6 \left( 1 - 2 t_1^2 t_2^2 + t_2^4 \right)=0,
\end{array}
\end{equation}
where for the sake of brevity we denoted~$t_a=\Im y^a$. Depending on the
value of the parameter~$\alpha$, the number of the solutions to the eqs.~(%
\ref{crit-axion-free}) and their stability change.
The stable solutions have all eigenvalues of the Hessian matrix positive,
while for the unstable ones~-- one of them is negative. In what follows we
consider only stable solutions.

\addtocounter{figure}{1} \newcounter{VD0D4} \setcounter{VD0D4}{%
\value{figure}} \addtocounter{figure}{1} Substituting stable solutions of~(%
\ref{crit-axion-free}) into~eq.~(\ref{SBH}) one gets the following behaviour
of the entropy with respect to the quantum parameter~(Fig.~\arabic{VD0D4}).
One can easily see that for~$\alpha > 2/(3\sqrt3)$ there are two solutions
to the attractor equations. The one having no classical limit is a 1/2-BPS
solution. Such an effect~-- i.e. the appearance of a BPS solution when the
quartic invariant~$I_4$ is negative was also observed in a quantum~$t^3$
model~\cite{BFMS1}.
\begin{tabular}{p{0.45\textwidth}p{0.45\textwidth}}
\begin{center}
\includegraphics[width=0.45\textwidth]{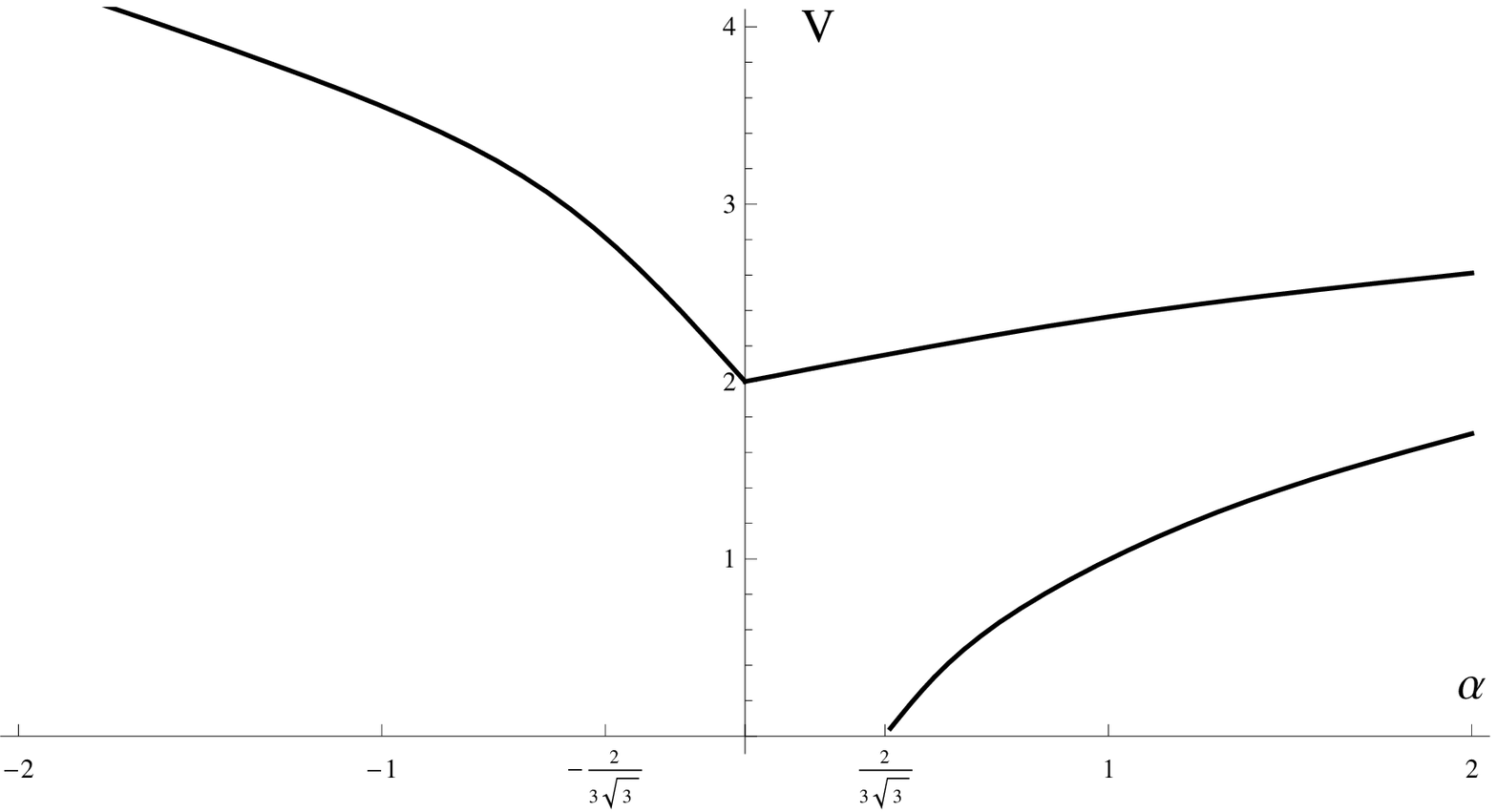}
Figure~\arabic{VD0D4}: Plot of~$v(y,\bar y)$ for~$D0-D4$ branes
\end{center}
&
\newcounter{WD0D4}
\setcounter{WD0D4}{\value{figure}}
\addtocounter{figure}{1}
\begin{center}
\includegraphics[width=0.5\textwidth]{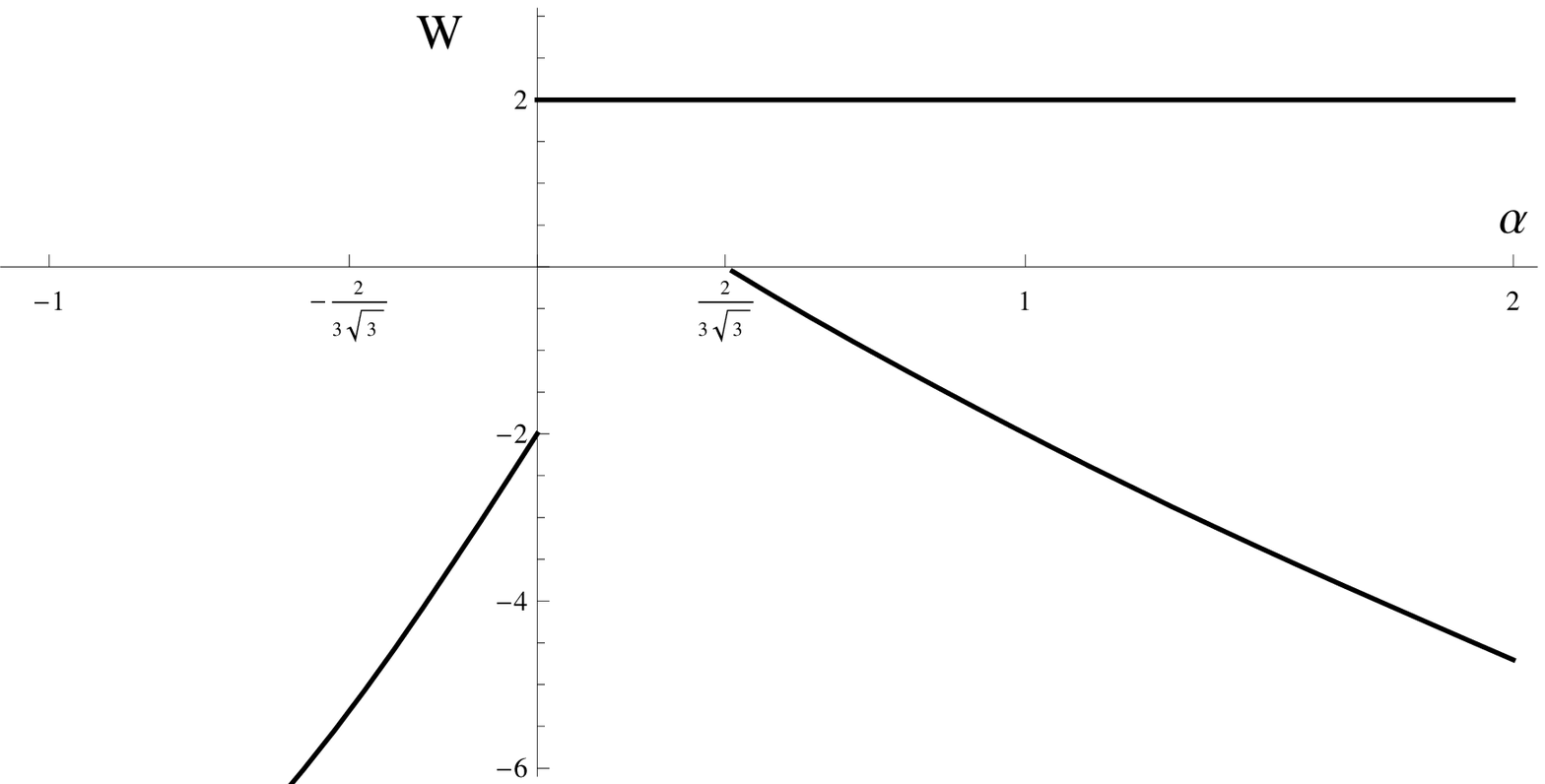}
Figure~\arabic{WD0D4}: Plot of~$W$ for~$D0-D4$ branes (in units of~$q_0$)
\end{center}
\end{tabular}
\par%
An interesting property is exhibited by the non-BPS solution with positive
value of the quantum parameter~$\alpha$~(Fig.~\arabic{WD0D4}). Being
evaluated on this solution, the superpotential does not depend on the
parameter~$\alpha$ and is equal to
\begin{equation*}
W=2q_0.
\end{equation*}
Generally, the superpotential without axion fields~(\ref{axion-free}) has
the form
\begin{equation}
W=q_0\left[ 1-2\Im y^1\Im y^2-(\Im y^2)^2\right] .
\end{equation}

\noindent By equating it to~$2q_0$ one obtains the following relation
between the moduli:
\begin{equation}
\Im y^1=-\frac{1+(\Im y^2)^2}{2\Im y^2}\,,
\end{equation}
which is consistent with the criticality condition~(\ref{VBHcrit}) provided
\begin{equation}
\alpha -\Im y^2+(\Im y^2)^3=0.  \label{cubicImy2}
\end{equation}
Obviously, such an algebraic equation has either one or three real solutions
depending on the value of~$\alpha$. When~$\alpha$ is positive, only one of
the solutions is stable. When it is negative, there is no stable solution
of~(\ref{cubicImy2}).

Just for the sake of mentioning it, we should say that there is another
solution yielding~$\alpha$-independent prepotential, namely~$W=0$, so that
the central charge vanishes as well. It holds provided
\begin{equation*}
D_t W = \lambda \frac{D_s W} {\left( \Im t \right)^3}
\end{equation*}
which is nothing but a ``quantum'' generalization of the zero central charge
condition found in~\cite{BMOS-1}. Since this solution turns out to be
unstable, we do not consider it further.

To conclude this section we consider a case when the quartic invariant~$I_4$
is positive. Classically, it is known to correspond to~$1/2$-BPS solutions
and thus in this case the black hole potential has no flat direction.
Although it is unlikely that a flat direction might appear when introducing
quantum corrections, we consider this case as well and list the results:

\begin{enumerate}
\item  there exists one~$1/2$-BPS solution, which is, naturally, stable~\cite
{FGK}. This solution pertains to the~$t^3+i\lambda$ model~\cite{BFMS1}.

\item  there exists a stable non-BPS~$Z=0$ solution with~$\Im y^1= \Im y^2$,
which does not have a classical limit.

\item  there exists a stable non-BPS~$Z=0$ solution, having its classical
limit as found in~\cite{BMOS-1}.

\item  there exist two unstable non-BPS solutions, having no classical
limit. They correspond to~$\alpha$-independent values of the superpotential:
either~$2q_0$ or zero.
\end{enumerate}

\newcounter{VBPSplot} \setcounter{VBPSplot}{\value{figure}}
\begin{multicols}{2}

The behaviour of the function~$v$ related to the black hole
potential via~(\ref{WVfactorized}) is presented
in~Fig.~\arabic{VBPSplot}. The properties of the three solutions
depicted here might be easily traced from the list above (remember
that only stable solutions are depicted).

To summarize, in the presence only of~$D0-D4$ branes the flat
direction of the classical black hole potential gets removed.
\begin{center}
\includegraphics[width=0.35\textwidth]{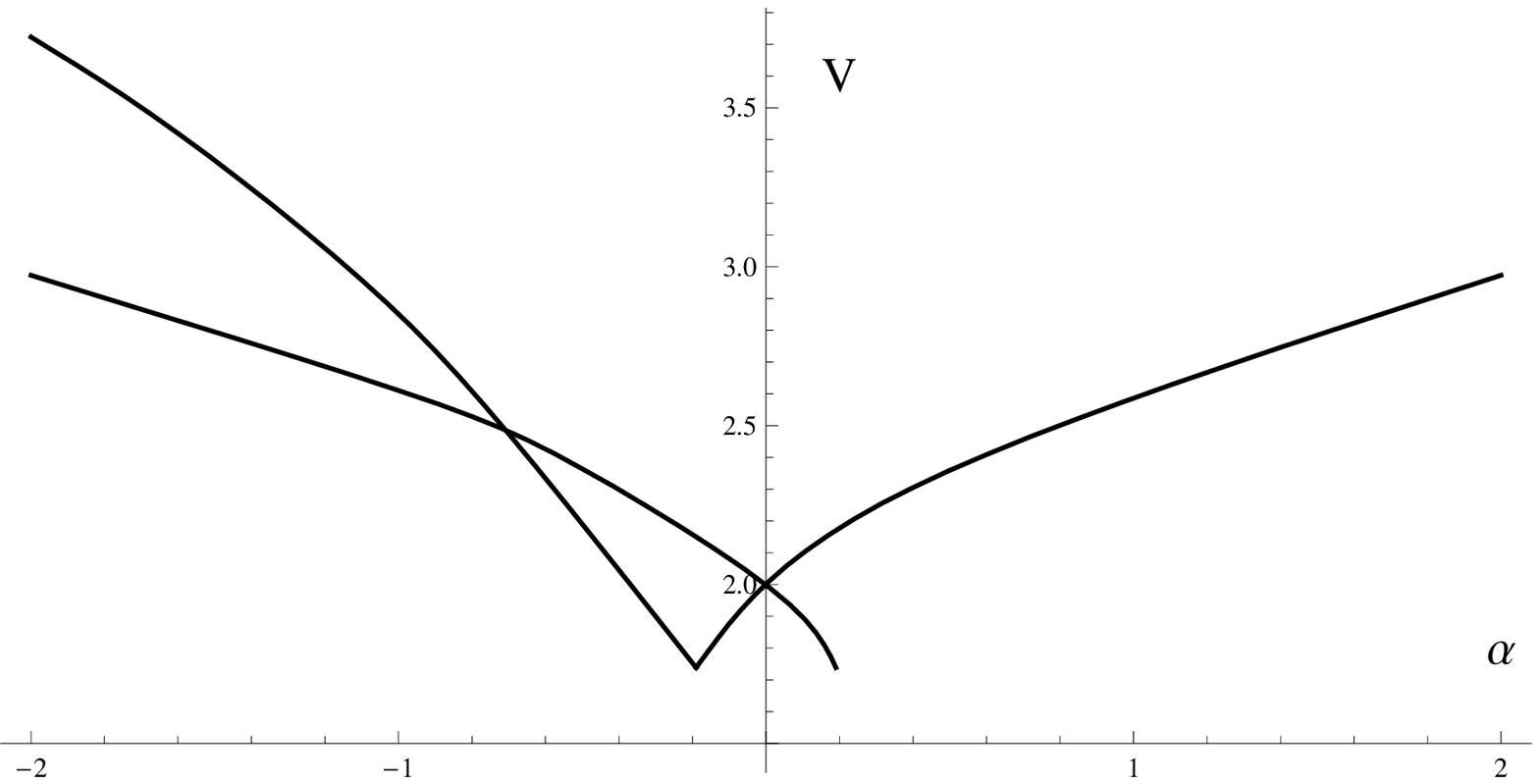}\\
Figure~\arabic{VBPSplot}: Plot of~$v(y,\bar y)$ for~$D0-D4$ branes,~$I_4>0$.
\end{center}
\end{multicols}
\noindent Unlike the classical case, there appear as well~$1/2$-BPS quantum
solutions with~$I_4<0$ and non-BPS ones with~$I_4>0$. This fact was observed
in~\cite{BMOS-1}. Another state not observed before is the~$\alpha$%
-independent one of the superpotential. A question to be yet clarified is a
correlation between the ground states of~BPS and non-BPS solutions. As one
sees from~Fig.\arabic{VD0D4} once both states~--BPS and non-BPS~-- are
present simultaneously for~$I_4<0$, the ground state energy of the BPS one
is lower than that of the non-BPS one. This is not valid any more for~$I_4>0$%
: from~Fig.\arabic{VBPSplot} one sees that there exists a value of~$\alpha$
when the BPS state has a lower energy than that of the non-BPS state, and
there exists as well a value of~$\alpha$ when the relation between energies
is opposite.

\section{$D2-D6$ branes}

Let us consider now a situation when only~$D2$ and~$D6$ branes are present.
The quartic invariant~$I_4$ in this case is equal to
\begin{equation}
I_4 = - p^0 q_1 q_2^2.
\end{equation}
Let us start with~$I_4>0$, when there exist classically~1/2-BPS and non-BPS~$%
Z=0$ attractors. The black hole potential exhibits no flat directions. In
the presence of~$D6$ branes, switching the quantum correction~$\alpha$ on,
the attractor eqs.~(\ref{VBHcrit}) are hard to be solved even numerically,
due to the presence of the~$D6$-brane charge~$p^0$~(\ref{WK}). Thus, the
question whether there is a quantum critical solution with positive~$I_4$
remains open.

Considering the case $I_4<0$, which classically admits only non-BPS
attractors, one can simplify the analysis by defining rescaled moduli~$y^a$
and a quantum parameter~$\alpha$ as follows:
\begin{equation}
s = \frac{y^1}{p^0\,q_1}\sqrt{\strut -I_4}\,,\quad t = \frac{y^2}{p^0\,q_2}%
\sqrt{\strut -I_4},\quad \lambda = \frac{\alpha}{(p^0)^2}\sqrt{\strut -I_4}.
\label{Sat-eve1}
\end{equation}
In the following treatment we choose~$p^0$ to be positive. Classically, the
black hole potential exhibits a non-compact flat direction
\begin{equation}  \label{flatDir-elec}
\Re y^1 = -\frac{\Re y^2}{1+(\Re y^2)^2},\qquad \Im y^1 = -\frac12\frac{%
1+(\Re y^2)^2}{1-(\Re y^2)^2},\qquad \Im y^2 = \pm \sqrt{\strut 1-(\Re y^2)^2%
},
\end{equation}
which spans a one-dimensional manifold~$SO(1,1)$~\cite{ferrara4}. The BH
potential evaluated on~eq.~(\ref{flatDir-elec}) gives as usual
\begin{equation}  \label{Sun-aft-1}
V_{BH} = \sqrt{-I_4} = \sqrt{p^0 q_1 q_2^2}.
\end{equation}

Introducing quantum effects destroys the classical flat direction~(\ref
{flatDir-elec}), and the critical solutions are reduced to a discrete set of
points (usually one or two) having the axion fields~$\Re y^a$ equal to zero.
The domain of positivity of the metric is defined by condition
\begin{equation}  \label{Sat-eve2}
\left[ \alpha -\Im y^1(\Im y^2)^2\right] \left[ \alpha +2\Im y^1(\Im y^2)^2%
\right] <0.
\end{equation}
\addtocounter{figure}{1}
\newcounter{VeffD2D6plot} \setcounter{VeffD2D6plot}{\value{figure}} %
\begin{multicols}{2}
In the considered case~$I_4<0$ with only~$D2$ and~$D6$ branes
present, the dependence of the minimum value of the quantum
corrected black hole potential~$V_{BH}$ on the quantum
parameter~$\alpha$ is presented in Fig.~\arabic{VeffD2D6plot}.
Numerical analysis does not support any evidence for the existence
of solutions with~\mbox{$|\alpha|>1$}, because such solutions fall
outside the domain of positivity defined by~eq.~(\ref{Sat-eve2}).
\begin{center}
\includegraphics[width=0.5\textwidth]{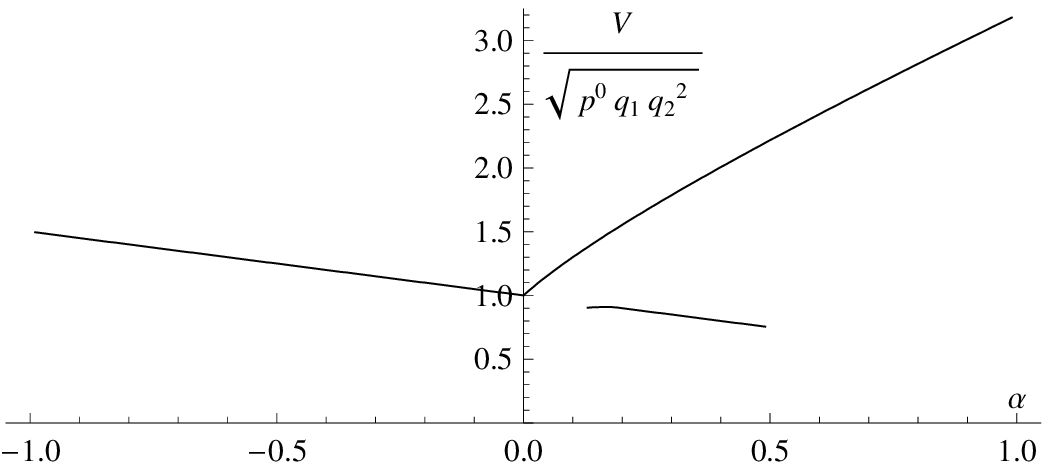}
\end{center}
Figure~\arabic{VeffD2D6plot}: Plot of~$V_{BH}/\sqrt{-I_4}$ for~$D2-D6$ branes and negative~$I_4$.

\end{multicols}
The plot for~$\alpha<0$ and the short curve for~$\alpha>0$ in~Fig.~%
\arabic{VeffD2D6plot} correspond to solutions to attractor equations with~$%
\alpha$-independent covariant derivatives of the superpotential
\begin{equation}
D_a W = q_a,\quad a=1,2.
\end{equation}

\section{$D0-D6$ branes}

\label{Sect4}

Let us now briefly analyze the~$D0-D6$ brane configuration in the~$st^2$
model. The corresponding charges~$p^0$ and~$q_0$ are those associated to the
Kaluza-Klein vector arising through dimensional reduction from~$d=5$ to~$d=4$%
~\cite{Ceresole}.

In this framework, the quartic invariant~$I_4$ is negative definite
\begin{equation}
I_4 = - (p^0 q_0)^2.
\end{equation}
In order to perform the analysis, it is once again convenient to introduce
rescaled moduli~$y^a$ and the quantum parameter~$\alpha$
\begin{equation}
s = \sqrt[3]{\frac{q_0}{p^0}}\,y^1, \qquad t = \sqrt[3]{\frac{q_0}{p^0}}%
\,y^2,\qquad \lambda = \frac{\alpha\, q_0}{2p^0}.
\end{equation}
The black hole potential has a flat direction
\begin{equation}  \label{Sat-eve3}
\Re y^1 = \Re y^2 = 0,\qquad \Im y^1 = \pm \frac1{(\Im y^2)^2},
\end{equation}
corresponding to non-BPS states. The sign plus is to be taken for~$p^0 q_0<0$
and the sign minus~-- otherwise. This flat direction is characterized by a
minimum of the black hole potential which turns out to be equal to
\begin{equation}
V_{BH} = \sqrt{-I_4} = \left| p^0\, q_0 \right|.
\end{equation}
Notice that, consistently with the analysis of~\cite{Ceresole}, the~$D0-D6$
configuration admits axion-free solutions at the classical level, and they
are actually general solutions.

Performing a thorough numerical analysis, an unexpected evidence emerges:
the classical non-BPS~$Z\neq 0$ flat direction of the~$st^2$ model in the~$%
D0-D6$ brane configuration seemingly survives the considered quantum
correction. This fact deeply distinguishes the~$D0-D6$ configuration from
the others treated above, when the quantum correction always lifts the flat
direction. Another feature of this quantum flat direction is the presence of
non-vanishing axion fields.

\begin{figure}[h]
\centering
\includegraphics[width=0.45\textwidth]{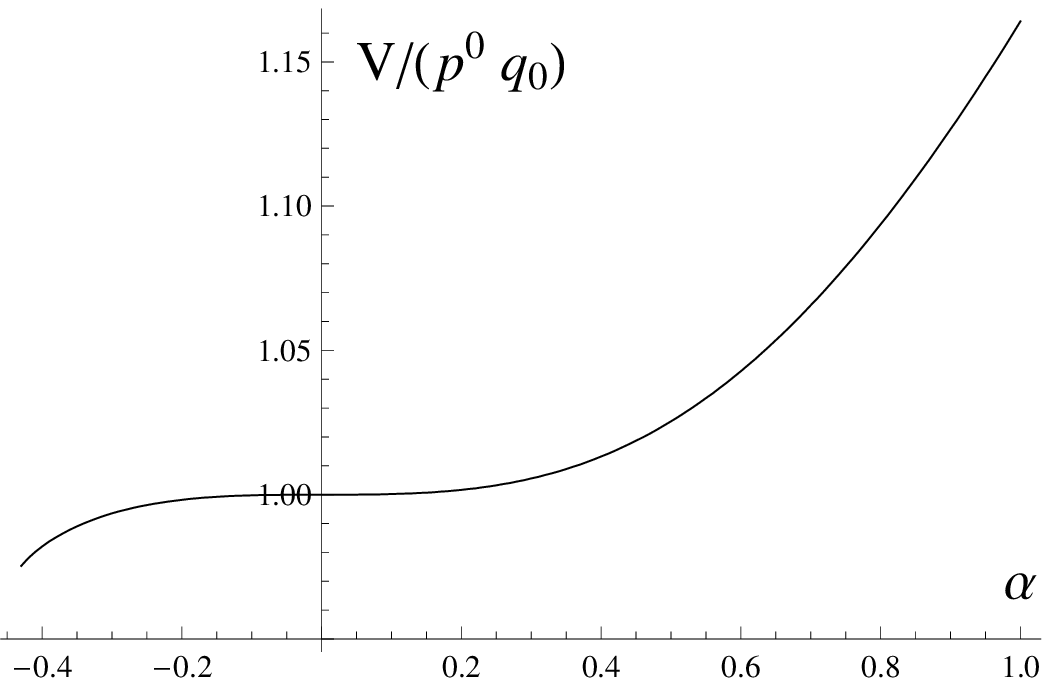}
\caption{Plot of the minimum of~$V_{BH}$ in the~$D0-D6$ brane configuration}
\label{VeffNeutral}
\end{figure}

Thus, one can conclude that the axion-free classical non-BPS~$Z\neq 0$ flat
direction is kept by the quantum corrections, but it gets distorted and
acquires non-zero values of the axion fields. Naturally, the black hole
potential takes a minimal value along the flat direction and the dependence
of this value on the quantum parameter is presented in~Fig.~\ref{VeffNeutral}%
.

\section{Conclusion and Outlook}

We addressed the issue of the fate of the unique non-BPS flat direction of~$%
st^2$ model in the presence of the most general class of quantum
perturbative corrections consistent with continuous axion-shift symmetry~
\cite{Peccei-Quinn}.

We performed our analysis in~$D0-D4$,~$D2-D6$ and~$D0-D6$ brane
configurations. For the first two cases we showed that the classical flat
direction gets lifted at the quantum level. The same behavior one may expect
for the unique non-BPS~$Z=0$ flat direction of the third element of the
cubic reducible sequence~$\frac{SU(1,1)}{U(1)}\times \frac{SO(2,n)}{%
SO(2)\times SO(n)}$ \cite{ferrara4}.

On the other hand, the analysis performed in the~$D0-D6$ brane configuration
yielded a somewhat surprising result: the classical flat direction gets
modified at the quantum level, acquiring a non-zero value of the axion
fields. The origin of such a deep difference among the brane configurations,
which is expected to hold in other models as well, are yet to be understood,
and we leave the study of this issue for future work.

Clearly, in the considered two moduli quantum corrected special K\"ahler~$d$%
-geometry based on a holomorphic prepotential, the phenomena of
``separation'' and ``transmutation'' of attractors, firstly observed in~\cite
{BFMS1}, also occur, with a richer case study, due to the presence of
non-BPS~$Z=0$ attractors.

By generalizing the results obtained in the present paper to the presence of
more than one flat direction, one would thus be led to state that only a few
classical attractors do remain attractors in a strict sense at the quantum
level. Consequently, at the quantum level the set of actual extremal black
hole attractors should be strongly constrained and reduced.

As a final remark, it is worth pointing out that in $N=8$ ($d=4$)
supergravity ``large'' $\frac{1}{8}$-BPS and non-BPS BHs exhibit $40$ and $%
42 $ flat directions, respectively \cite{ADF-Duality-d=4,Ferrara-Marrani-1}.
If $N=8$ supergravity is a finite theory of quantum gravity (see \textit{e.g.%
} \cite{Bern} and Refs. therein), it would be interesting to
understand whether these flat directions may be removed at all by
perturbative and/or non-perturbative quantum effects.

\section*{\textbf{Acknowledgments}}

A. M. would like to thank the Department of Physics, Theory Unit Group at
CERN, and the Berkeley Center for Theoretical Physics (CTP) of the
University of California, Berkeley, USA, where part of this work was done,
for kind hospitality and stimulating environment.


The work of S.F. has also been supported in part by D.O.E. grant
DE-FG03-91ER40662, Task C, and by the Miller Institute for Basic Research in
Science, University of California, Berkeley, CA, USA.

The work of A. M. has been supported by an INFN visiting Theoretical
Fellowship at SITP, Stanford University, Stanford, CA, USA.

The work of A.S. has been supported by a Junior Grant of the \textit{%
``Enrico Fermi''} Center, Rome, in association with INFN Frascati National
Laboratories.


\begin{thebibliography}{99}
\bibitem{FKS}  S. Ferrara, R. Kallosh and A. Strominger, $N=2$\textit{\
Extremal Black Holes}, Phys. Rev.~\textbf{D52}, 5412 (1995), \texttt{%
hep-th/9508072}.

\bibitem{Strom}  A. Strominger, \textit{Macroscopic Entropy of }$N=2$\textit{%
\ Extremal Black Holes}, Phys. Lett.~\textbf{B383}, 39 (1996), \texttt{%
hep-th/9602111}.

\bibitem{FK1}  S. Ferrara and R. Kallosh, \textit{Supersymmetry and
Attractors}, Phys. Rev.~\textbf{D54}, 1514 (1996), \texttt{hep-th/9602136}.

\bibitem{FK2}  S. Ferrara and R. Kallosh, \textit{Universality of
Supersymmetric Attractors}, Phys. Rev.~\textbf{D54}, 1525 (1996), \texttt{%
hep-th/9603090}.

\bibitem{FGK}  S. Ferrara, G. W. Gibbons and R. Kallosh, \textit{Black Holes
and Critical Points in Moduli Space}, Nucl. Phys.~\textbf{B500}, 75 (1997),
\texttt{hep-th/9702103}.

\bibitem{bellucci2}  S. Bellucci, S. Ferrara, R. Kallosh and A. Marrani,
\textit{Extremal Black Hole and Flux Vacua Attractors}, contribution to the
Proceedings of the Winter School on Attractor Mechanism 2006 (SAM2006),
20-24 March 2006, INFN-LNF, Frascati, Italy, \texttt{arXiv:0711.4547}.

\bibitem{BPS}  G. W. Gibbons and C. M. Hull, \textit{A Bogomol'ny Bound for
General Relativity and Solitons in }$N=2$\textit{\ Supergravity}, Phys. Lett~%
\textbf{B109}, 190 (1982).

\bibitem{Ferrara-Marrani-1}  S. Ferrara and A. Marrani, $N=8$\textit{\
non-BPS Attractors, Fixed Scalars and Magic Supergravities}, Nucl. Phys.~%
\textbf{B788}, 63 (2008), \texttt{arXiV:0705.3866}.

\bibitem{ferrara4}  S. Ferrara and A. Marrani, \textit{On the Moduli Space
of non-BPS Attractors for~}$N=2$\textit{\ Symmetric Manifolds}, Phys. Lett.~%
\textbf{B652}, 111 (2007), \texttt{arXiV:0706.1667}.

\bibitem{bellucci1}  S. Bellucci, S. Ferrara, M. Günaydin and A.
Marrani, \textit{Charge Orbits of Symmetric Special Geometries and Attractors%
}, Int. J. Mod. Phys~\textbf{A21}, 5043 (2006), \texttt{hep-th/0606209}.

\bibitem{BFMS1}  S. Bellucci, S. Ferrara, A. Marrani and A. Shcherbakov,
\textit{Splitting of Attractors in }$\mathit{1}$\textit{-modulus Quantum
Corrected Special Geometry}, JHEP~\textbf{0802}, 088 (2008), \texttt{%
arXiv:0710.3559}.

\bibitem{Shmakova}  M. Shmakova, \textit{Calabi-Yau black holes}, Phys. Rev.~%
\textbf{D56}, 540 (1997), \texttt{hep-th/9612076}.

\bibitem{Saraikin-Vafa-1}  K. Saraikin and C. Vafa, \textit{%
Non-supersymmetric Black Holes and Topological Strings}, Class. Quant. Grav.~%
\textbf{25}, 095007 (2008), \texttt{hep-th/0703214}.

\bibitem{stu-unveiled}  S.~Bellucci, S.~Ferrara, A.~Marrani and A.~Yeranyan,
\textit{stu Black Holes Unveiled}, Entropy~\textbf{10(4)}, 507-555 (2008),
\texttt{arXiV:0807.3503}.

\bibitem{BMOS-1}  S. Bellucci, A. Marrani, E. Orazi and A. Shcherbakov,
\textit{Attractors with Vanishing Central Charge}, Phys. Lett.~\textbf{B655}%
, 185 (2007), \texttt{arXiV:0707.2730}.

\bibitem{Peccei-Quinn}  R. D.~Peccei and H. R.~Quinn, \textit{Constraints
imposed by CP conservation in the presence of instantons}, Phys. Rev.~%
\textbf{D16}, 1791 (1977); R. D.~Peccei and H. R.~Quinn, \textit{CP
conservation in the presence of instantons}, Phys. Rev. Lett.~\textbf{38},
1440 (1977); R. D.~Peccei and H. R.~Quinn, \textit{Some aspects of instantons%
}, Nuovo Cim.~\textbf{A41}, 309 (1977).

\bibitem{CFG}  S. Cecotti, S. Ferrara and L. Girardello, \textit{Geometry of
Type }$\mathit{II}$\textit{\ Superstrings and the Moduli of Superconformal
Field Theories}, Int. J. Mod. Phys.~\textbf{A4}, 2475 (1989).

\bibitem{G}  A. Giryavets, \textit{New Attractors and Area Codes}, JHEP~%
\textbf{0603}, 020 (2006), \texttt{hep-th/0511215}.

\bibitem{CDLOGP1}  P. Candelas, X. C. De La Ossa, P. S. Green and L. Parkes,
\textit{A Pair of Calabi-Yau Manifolds as an Exactly Soluble Superconformal
Theory}, Nucl. Phys. \textbf{B359}, 21 (1991); P. Candelas, X. C. De La
Ossa, P. S. Green and L. Parkes, \textit{An Exactly Soluble Superconformal
Theory from a Mirror Pair of Calabi-Yau Manifolds}, Phys. Lett. \textbf{B258}%
, 118 (1991).

\bibitem{HKTY}  S. Hosono, A. Klemm, S. Theisen and Shing-Tung Yau, \textit{%
Mirror symmetry, mirror map and applications to Calabi-Yau hypersurfaces},
Commun. Math. Phys. \textbf{167}, 301 (1995), \texttt{hep-th/9308122}.

\bibitem{Quantum-N=2}  K. Behrndt, G. Lopes Cardoso, B. de Wit, R. Kallosh,
D. Lüst and T. Mohaupt, \textit{Classical and quantum }$N\mathit{=2}$%
\textit{\ supersymmetric black holes}, Nucl. Phys. \textbf{B488}, 236
(1997), \texttt{hep-th/9610105}.

\bibitem{Alvarez-Gaume}  L. Alvarez-Gaume, D. Z. Freedman, \textit{%
Geometrical Structure and Ultraviolet Finiteness in the Supersymmetric Sigma
Model}, Commun. Math. Phys. \textbf{80}, 443 (1981).

\bibitem{Grisaru}  M. T. Grisaru, A. van de Ven and D. Zanon, \textit{Four
Loop Beta Function for the }$N\mathit{=1}$\textit{\ and }$N\mathit{=2}$%
\textit{\ Supersymmetric Nonlinear Sigma Model in Two Dimensions}, Phys.
Lett. \textbf{B173}, 423 (1986); M. T. Grisaru, A. van de Ven and D. Zanon,
\textit{Two Dimensional Supersymmetric Sigma Models on Ricci Flat Kähler
Manifolds are not Finite}, Nucl. Phys. \textbf{B277}, 388 (1986); M. T.
Grisaru, A. van de Ven and D. Zanon, \textit{Four Loop Divergences for the }$%
N\mathit{=1}$\textit{\ Supersymmetric Nonlinear Sigma Model in Two Dimensions%
}, Nucl. Phys. \textbf{B277}, 409 (1986).

\bibitem{Ceresole}  A. Ceresole, S. Ferrara and A. Marrani, $\mathit{4d}$%
\textit{/}$\mathit{5d}$ \textit{Correspondence for the Black Hole Potential
and its Critical Points}, Class. Quant. Grav.~\textbf{24}, 5651 (2007),
\texttt{arXiV:0707.0964}.

\bibitem{ADF-Duality-d=4}  L. Andrianopoli, R. D'Auria and S. Ferrara, $%
\mathit{U}$\textit{\ invariants, black hole entropy and fixed scalars},
Phys. Lett. \textbf{B403}, 12 (1997), \texttt{hep-th/9703156}.

\bibitem{Bern}  Z. Bern, J. J. Carrasco, L. J. Dixon, H. Johansson, D. A.
Kosower and R. Roiban, \textit{Three-Loop Finiteness of }$N=8$\textit{\
Supergravity}, Phys. Rev. Lett. \textbf{98}, 161303 (2007), \texttt{%
hep-th/0702112}.
\end{thebibliography}
\end{document}